

\def\pmb#1{\setbox0=\hbox{#1}%
  \hbox{\kern-.025em\copy0\kern-\wd0
  \kern.05em\copy0\kern-\wd0
  \kern-0.025em\raise.0433em\box0} }

\catcode`@=11
\def\leftrightarrowfill{$\m@th\mathord\leftarrow \mkern-6mu
  \cleaders\hbox{$\mkern-2mu \mathord- \mkern-2mu$}\hfill
  \mkern-6mu \mathord\rightarrow$}
\def\overleftrightarrow#1{\vbox{\ialign{##\crcr
     \leftrightarrowfill\crcr\noalign{\kern-1pt\nointerlineskip}
     $\hfil\displaystyle{#1}\hfil$\crcr}}}
\catcode`@=12

\def\approxge{\hbox {\hfil\raise .4ex\hbox{$>$}\kern-.75 em
\lower .7ex\hbox{$\sim$}\hfil}}
\def\approxle{\hbox {\hfil\raise .4ex\hbox{$<$}\kern-.75 em
\lower .7ex\hbox{$\sim$}\hfil}}

\def \abstract#1 {\vskip 0.5truecm\sepline\vskip 0.5truecm
$$\vbox{\hsize=15truecm\noindent #1}$$}
\def \SISSA#1#2 {\vfil\vfil\centerline{Ref. S.I.S.S.A. #1 CM (#2)}}
\def \PACS#1 {\vfil\line{\hfil\hbox to 15truecm{PACS numbers: #1 \hfil}\hfil}}

\def \hfigure
     #1#2#3       {\midinsert \vskip #2 truecm $$\vbox{\hsize=14.5truecm
             \seven\baselineskip=10pt\noindent {\bcp \noindent Figure  #1}.
                   #3 } $$ \vskip -20pt \endinsert }

\def \hfiglin
     #1#2#3       {\midinsert \vskip #2 truecm $$\vbox{\hsize=14.5truecm
              \seven\baselineskip=10pt\noindent {\bcp \hfil\noindent
                   Figure  #1}. #3 \hfil} $$ \vskip -20pt \endinsert }

\def \vfigure
     #1#2#3#4     {\dimen0=\hsize \advance\dimen0 by -#3truecm
                   \midinsert \vbox to #2truecm{ \seven
                   \parshape=1 #3truecm \dimen0 \baselineskip=10pt \vfill
                   \noindent{\bcp Figure #1} \pretolerance=6500#4 \vfill }
                   \endinsert }

%
\def \ref
     #1#2         {\smallskip \item{[#1]}#2}
\def \sepline     {\medskip\centerline{\vbox{\hrule width5truecm}} \medskip}

\def \tabrul2     {\noalign{\vskip 5truept \hrule \vskip 2truept \hrule
                   \vskip 5truept} }


\footline={\ifnum\pageno>0 \tenrm \hss \folio \hss \fi }

\def\today
 {\count10=\year\advance\count10 by -1900 \number\day--\ifcase
  \month \or Jan\or Feb\or Mar\or Apr\or May\or Jun\or
             Jul\or Aug\or Sep\or Oct\or Nov\or Dec\fi--\number\count10}

\def\hour{\count10=\time\count11=\count10
\divide\count10 by 60 \count12=\count10
\multiply\count12 by 60 \advance\count11 by -\count12\count12=0
\number\count10 :\ifnum\count11 < 10 \number\count12\fi\number\count11}

\def\draft{
   \baselineskip=20pt
   \def\makeheadline{\vbox to 10pt{\vskip-22.5pt
   \line{\vbox to 8.5pt{}\the\headline}\vss}\nointerlineskip}
   \headline={\hfill \seven {\bcp Draft version}: today is \today\ at \hour
              \hfill}
          }

%
%

%
\catcode`@=11
%
%
\def\b@lank{ }

\newif\if@simboli
\newif\if@riferimenti

\newwrite\file@simboli
\def\simboli{
    \immediate\write16{ !!! Genera il file \jobname.SMB }
    \@simbolitrue\immediate\openout\file@simboli=\jobname.smb}

\newwrite\file@ausiliario
\def\riferimentifuturi{
    \immediate\write16{ !!! Genera il file \jobname.AUX }
    \@riferimentitrue\openin1 \jobname.aux
    \ifeof1\relax\else\closein1\relax\input\jobname.aux\fi
    \immediate\openout\file@ausiliario=\jobname.aux}

\newcount\eq@num\global\eq@num=0
\newcount\sect@num\global\sect@num=0

\newif\if@ndoppia
\def\numerazionedoppia{\@ndoppiatrue\gdef\la@sezionecorrente{\the\sect@num}}

\def\se@indefinito#1{\expandafter\ifx\csname#1\endcsname\relax}
\def\spo@glia#1>{} 

\newif\if@primasezione
\@primasezionetrue

\def\s@ection#1\par{\immediate
    \write16{#1}\if@primasezione\global\@primasezionefalse\else\goodbreak
    \vskip\spaziosoprasez\fi\noindent
    {\bf#1}\nobreak\vskip\spaziosottosez\nobreak\noindent}
%

\def\sezpreset#1{\global\sect@num=#1
    \immediate\write16{ !!! sez-preset = #1 }   }

\def\spaziosoprasez{50pt plus 60pt}
\def\spaziosottosez{15pt}

\def\sref#1{\se@indefinito{@s@#1}\immediate\write16{ ??? \string\sref{#1}
    non definita !!!}
    \expandafter\xdef\csname @s@#1\endcsname{??}\fi\csname @s@#1\endcsname}

\def\autosez#1#2\par{
    \global\advance\sect@num by 1\if@ndoppia\global\eq@num=0\fi
    \xdef\la@sezionecorrente{\the\sect@num}
    \def\usa@getta{1}\se@indefinito{@s@#1}\def\usa@getta{2}\fi
    \expandafter\ifx\csname @s@#1\endcsname\la@sezionecorrente\def
    \usa@getta{2}\fi
    \ifodd\usa@getta\immediate\write16
      { ??? possibili riferimenti errati a \string\sref{#1} !!!}\fi
    \expandafter\xdef\csname @s@#1\endcsname{\la@sezionecorrente}
    \immediate\write16{\la@sezionecorrente. #2}
    \if@simboli
      \immediate\write\file@simboli{ }\immediate\write\file@simboli{ }
      \immediate\write\file@simboli{  Sezione
                                  \la@sezionecorrente :   sref.   #1}
      \immediate\write\file@simboli{ } \fi
    \if@riferimenti
      \immediate\write\file@ausiliario{\string\expandafter\string\edef
      \string\csname\b@lank @s@#1\string\endcsname{\la@sezionecorrente}}\fi
    \goodbreak\vskip 48pt plus 60pt
\centerline{\lltitle #2}                     
\par\nobreak\vskip 15pt \nobreak\noindent}

\def\semiautosez#1#2\par{
    \gdef\la@sezionecorrente{#1}\if@ndoppia\global\eq@num=0\fi
    \if@simboli
      \immediate\write\file@simboli{ }\immediate\write\file@simboli{ }
      \immediate\write\file@simboli{  Sezione ** : sref.
          \expandafter\spo@glia\meaning\la@sezionecorrente}
      \immediate\write\file@simboli{ }\fi
\noindent\lltitle \s@ection#2 \par}


\def\eqpreset#1{\global\eq@num=#1
     \immediate\write16{ !!! eq-preset = #1 }     }

\def\eqref#1{\se@indefinito{@eq@#1}
    \immediate\write16{ ??? \string\eqref{#1} non definita !!!}
    \expandafter\xdef\csname @eq@#1\endcsname{??}
    \fi\csname @eq@#1\endcsname}

\def\eqlabel#1{\global\advance\eq@num by 1
    \if@ndoppia\xdef\il@numero{\la@sezionecorrente.\the\eq@num}
       \else\xdef\il@numero{\the\eq@num}\fi
    \def\usa@getta{1}\se@indefinito{@eq@#1}\def\usa@getta{2}\fi
    \expandafter\ifx\csname @eq@#1\endcsname\il@numero\def\usa@getta{2}\fi
    \ifodd\usa@getta\immediate\write16
       { ??? possibili riferimenti errati a \string\eqref{#1} !!!}\fi
    \expandafter\xdef\csname @eq@#1\endcsname{\il@numero}
    \if@ndoppia
       \def\usa@getta{\expandafter\spo@glia\meaning
       \la@sezionecorrente.\the\eq@num}
       \else\def\usa@getta{\the\eq@num}\fi
    \if@simboli
       \immediate\write\file@simboli{  Equazione
            \usa@getta :  eqref.   #1}\fi
    \if@riferimenti
       \immediate\write\file@ausiliario{\string\expandafter\string\edef
       \string\csname\b@lank @eq@#1\string\endcsname{\usa@getta}}\fi}

\def\autoeqno#1{\eqlabel{#1}\eqno(\csname @eq@#1\endcsname)}
\def\autoleqno#1{\eqlabel{#1}\leqno(\csname @eq@#1\endcsname)}
\def\eqrefp#1{(\eqref{#1})}


\def\eq{\autoeqno}
\def\req{\eqrefp}
\def\chap{\autosez}        



\newcount\cit@num\global\cit@num=0

\newwrite\file@bibliografia
\newif\if@bibliografia
\@bibliografiafalse

\def\lp@cite{[}
\def\rp@cite{]}
\def\trap@cite#1{\lp@cite #1\rp@cite}
\def\lp@bibl{[}
\def\rp@bibl{]}
\def\trap@bibl#1{\lp@bibl #1\rp@bibl}

\def\refe@renza#1{\if@bibliografia\immediate        
    \write\file@bibliografia{
    \string\item{\trap@bibl{\cref{#1}}}\string
    \bibl@ref{#1}\string\bibl@skip}\fi}

\def\ref@ridefinita#1{\if@bibliografia\immediate\write\file@bibliografia{
    \string\item{?? \trap@bibl{\cref{#1}}} ??? tentativo di ridefinire la
      citazione #1 !!! \string\bibl@skip}\fi}

\def\bibl@ref#1{\se@indefinito{@ref@#1}\immediate
    \write16{ ??? biblitem #1 indefinito !!!}\expandafter\xdef
    \csname @ref@#1\endcsname{ ??}\fi\csname @ref@#1\endcsname}

\def\c@label#1{\global\advance\cit@num by 1\xdef            
   \la@citazione{\the\cit@num}\expandafter
   \xdef\csname @c@#1\endcsname{\la@citazione}}

\def\bibl@skip{\vskip +4truept}


\def\stileincite#1#2{\global\def\lp@cite{#1}\global   
    \def\rp@cite{#2}}                                 
\def\stileinbibl#1#2{\global\def\lp@bibl{#1}\global   
    \def\rp@bibl{#2}}                                 

\def\citpreset#1{\global\cit@num=#1
    \immediate\write16{ !!! cit-preset = #1 }    }

\def\autobibliografia{\global\@bibliografiatrue\immediate
    \write16{ !!! Genera il file \jobname.BIB}\immediate
    \openout\file@bibliografia=\jobname.bib}

\def\cref#1{\se@indefinito                  
   {@c@#1}\c@label{#1}\refe@renza{#1}\fi\csname @c@#1\endcsname}

\def\cite#1{\trap@cite{\cref{#1}}}                  
\def\ccite#1#2{\trap@cite{\cref{#1},\cref{#2}}}     
\def\ncite#1#2{\trap@cite{\cref{#1}--\cref{#2}}}    
\def\upcite#1{$^{\,\trap@cite{\cref{#1}}}$}               
\def\upccite#1#2{$^{\,\trap@cite{\cref{#1},\cref{#2}}}$}  
\def\upncite#1#2{$^{\,\trap@cite{\cref{#1}-\cref{#2}}}$}  

\def\clabel#1{\se@indefinito{@c@#1}\c@label           
    {#1}\refe@renza{#1}\else\c@label{#1}\ref@ridefinita{#1}\fi}

\def\biblskip#1{\def\bibl@skip{\vskip #1}}           

\def\insertbibliografia{\if@bibliografia             
    \immediate\write\file@bibliografia{ }
    \immediate\closeout\file@bibliografia
    \catcode`@=11\input\jobname.bib\catcode`@=12\fi}


\def\commento#1{\relax}
\def\biblitem#1#2\par{\expandafter\xdef\csname @ref@#1\endcsname{#2}}


\catcode`@=12


\magnification=1200
\topskip 20pt
\def\interlinea{\baselineskip=16pt}
\def\standardpage{\vsize=20.7truecm\voffset=+1.truecm
                  \hsize=15.truecm\hoffset=+10truemm
                  \parindent=1.2truecm}

\tolerance 100000
\biblskip{+8truept}                        
\def\hbup{\hfill\break\baselineskip 16pt}  


\global\newcount\notenumber \global\notenumber=0
\def\note #1 {\global\advance\notenumber by1 \baselineskip 10pt
              \footnote{$^{\the\notenumber}$}{\nine #1} \interlinea}



\font\text=cmr10
\font\scal=cmsy5
\font\it=cmti10

\font\title=cmbx10 scaled \magstep3      
\font\ltitle=cmbx12 scaled \magstep1
\font\lltitle=cmbx12 scaled \magstep1

\font\abs=cmti10 scaled \magstep1        

\font\seven=cmr7                         
\font\nine=cmr9                         
\font\bcp=cmbx7








\def\gtrsim{\ \rlap{\raise 2pt \hbox{$>$}}{\lower 2pt \hbox{$\sim$}}\ }
\def\lesssim{\ \rlap{\raise 2pt \hbox{$<$}}{\lower 2pt \hbox{$\sim$}}\ }


\def\mn{\medskip\noindent}
\def\bs{\bigskip}
\def\hb{\hfil\break}

\def\scss{\scriptscriptstyle}

\def\o{\over}



\def\ea{{\it et.al.}}
\def\ib{{\it ibid.\ }}

\def\npb#1{Nucl. Phys. {\bf B#1},}
\def\plb#1{Phys. Lett. {\bf B#1},}
\def\prd#1{Phys. Rev. {\bf D#1},}
\def\prl#1{Phys. Rev. Lett. {\bf #1},}
\def\zpc#1{Z. Phys. {\bf C#1},}
\def\prep#1{Phys. Rep. {\bf #1},}


\stileincite{}{}     
\numerazionedoppia   

\interlinea
\standardpage
\text                


\def\scss#1{{\scriptscriptstyle #1}}

\def\o{\over}



\def\G{{\cal G_{\rm SM}}}
\def\E{{\rm E}_6}

\def\N{{\hbox{\scal N}}}
\def\K{{\hbox{\scal K}}}

\def\pr{\prime}


\font\mbf=cmmib10  scaled \magstep1      

\def\bfmu{{\hbox{\mbf\char'026}}}

\def\bfe{{\hbox{\mbf\char'145}}}



\autobibliografia
\pageno=0
\vsize=23.8truecm
\hsize=15.7truecm
\voffset=-1.truecm
\hoffset=+6truemm
\baselineskip 14pt
\rightline{UM-TH 92--19}\par\noindent

\bs\bs\bs
\centerline{\ltitle   Z$^{\displaystyle\bf \prime}$,
new fermions and flavor changing processes.}
\medskip
\medskip
\centerline{\ltitle
   Constraints on E$_{\, \displaystyle\bf 6}\,$ models from
\bfmu $\ \longrightarrow$ \bfe\bfe\bfe.}
\bs\bs
\bs\bs                                   
                 \centerline{Enrico Nardi}
\bs
\centerline{\it Randall Laboratory of Physics, University of Michigan,
           Ann Arbor, MI 48109--1120}
\vskip 1truecm                             

\medskip
\centerline  { \bf {\abs Abstract}}      
\bs

\noindent
We study a new class of flavor changing interactions, which can arise
in models based on extended gauge groups (rank $>$4) when new charged
fermions are present together with a new neutral gauge boson. We
discuss the cases in which the flavor changing couplings in the new
neutral current coupled to the $Z^\prime$ are theoretically expected
to be large, implying that the observed suppression of neutral flavor
changing transitions must be provided by heavy $Z^\prime$ masses
together with small $Z$-$Z^\prime$ mixing angles. Concentrating on
E$_6$ models, we show how the tight experimental limit on $\mu
\rightarrow eee$ implies serious constraints on the $Z^\prime$ mass
and mixing angle. We conclude that if the value of the flavor changing
parameters is assumed to lie in a theoretically natural range, in most
cases the presence of a $Z^\prime$ much lighter than 1 TeV is
unlikely.
\bs
\noindent
PACS number(s): 13.10.+q,12.10.Dm,12.15.Cc,14.60.Jj
\vfill
\noindent
--------------------------------------------\phantom{-} \hb
\leftline{E-mail: nardi@umiphys.bitnet}
\bigskip
\leftline{UM-TH 92--19}
                   \bigskip
\centerline{August 1992}

\bs\bs
\eject

\standardpage                            
\interlinea                              
\null

\chap{Int} I. Introduction

The large set of accurate measurements performed during the last few
years has established that the standard electroweak theory provides an
excellent description of the particle physics phenomena up to about
100 GeV. It should also be stressed that the present theory
accommodates in a satisfactory way the whole spectrum of known
particles, and only two states, the top--quark and the Higgs scalar
that are necessary for the consistency of the model, have not been
discovered yet.
Another feature of the standard model (SM) that explains in a
satisfactory way a large set of experimental limits on rare processes,
is the Glashow-Iliopoulos-Maiani 
suppression of flavor changing
neutral currents (FCNC) in the quark sector, together with the absence
of lepton flavor violating (LFV) currents.
All these features are quite peculiar of the SM, and in general most
of its possible extensions
predict a larger spectrum of states as well as larger rates for
FCNC processes.
Clearly the direct detection of any new unconventional particle would
be a major breakthrough towards the identification of a
new and more fundamental theory, but at the same time it is not
unlikely that some hints on the existence of new physics will come
from the observation of rare processes at rates larger than
what expected in the standard theory.

The aim of this paper is to analyze a new class of
FCNC interactions that are generally
present in most of the extensions of the
SM which predict one (or more) additional neutral gauge boson
$Z_1$ together with
new charged fermions, and that are induced by $Z_1$ interactions.
In general the $Z_1$ is expected to be mixed with the standard
$Z_0$, and as a consequence both the resulting mass eigenstates,
that we will denote  as $Z$ and $Z^\pr$, will have flavor
changing couplings to the known charged fermions.

In particular we will concentrate on
LFV interactions, which are strictly forbidden in the SM,
and we will show that they could provide a clear signature
for this kind of new physics. In turn, the existing limits on LFV processes
imply serious constraints for several interesting models.
To stress the power of these constraints we will apply them to
a class of E$_6$ grand unified theories (GUTs),
which are a well known example of
theories where additional fermions and new neutral gauge bosons are
simultaneously present, and provide a good
frame for illustrating the kind of effects that we
want to explore. In an attempt of quantifying the constraints, we
will assume for the relevant E$_6$ LFV couplings
a range of values that we believe is theoretically natural,
and we will then turn the existing limits on LFV decays into
limits on the $Z^\prime$ parameters.
Though model dependent, our bounds turn out to be
much tighter than the present
limits obtained from direct searches at colliders [\cite{zp-direct}]
or as derived from the analysis of other $Z^\prime$ indirect effects
[\cite{fit6}\clabel{l-luo}--\cite{zp-new}].

The interesting possibility of violating
the conservation of lepton family number
as a consequence of $Z_1$
interactions was already briefly analyzed in [\cite{zp-fc}]
for a general class of extended electroweak models.
It was shown that LFV decays of the standard neutral gauge boson
like $Z  \rightarrow \ell_i \ell_j$, ($i\ne j$,
hereafter understood) can be induced by the presence of a $Z_1$ \
$i)$ if a sizeable mixing with the $Z_1$ is present, and \
$ii)$ if the new neutral gauge boson do not couple universally to the
fermion generations.
While the first condition is a natural feature of extended gauge models,
it could seem that the second one is somewhat more difficult to realize.
However, we will show that both these conditions are naturally satisfied
in extended gauge models that, like E$_6$, predict also new charged
fermions. This was already noted in a recent paper [\cite{fit6}],
where the consequences of the simultaneous presence of new neutral gauge bosons
and new fermions were analyzed, and a general formalism for taking
into account their combined effects was outlined.

In models like E$_6$, where several new neutral leptons are present
and a quite general form for the corresponding mass matrices and mixing
patterns is possible, LFV processes like $Z \rightarrow \ell_i
\ell_j$ arise naturally due to the contributions of loop
diagrams [\cite{e6-fcloop}]. A $Z_0\ell_i \ell_j$ vertex for
the gauge eigenstate $Z_0$ boson can however
appear already at the tree level, due to the presence of new exotic
charged leptons that could be mixed with the standard ones. A general
discussion of this kind of flavor changing vertices is given
e.g. in [\cite{ll1}], while the particular case of E$_6$ models is
analyzed in [\cite{e6-fcrizzo}].
The interest of considering flavor changing $Z_1$ interactions
stems from the fact that while in general the $Z_0$ flavor changing vertices
are strongly suppressed as the ratio between the masses of the known
and of the knew fermions, in some class of models no suppression factors are
expected for the $Z_1 f_i f_j$ vertices.
Then, in the context of these models, the
new interactions due to the additional gauge boson
could well represent the main source of flavor changing transitions.

In Sec. II
we will first review the essential formalism for dealing with
gauge boson and fermion mixing effects [\cite{fit6}] and
we will also
discuss the theoretical expectations  for the various flavor changing
parameters.
In Sec. III we will concentrate on E$_6$ models. We will confront the
theoretical expression for the decay
$\mu \rightarrow eee$ with the extremely stringent experimental limit
$Br(\mu^+ \rightarrow e^+e^+e^-) < 1\cdot 10^{-12}$
obtained by the SINDRUM collaboration [\cite{sindrum}],
and we will derive constraints on the mass of the new gauge boson
and on the
$Z_0$--$Z_1$ mixing angle as a function of the LFV parameters.
Finally, in Sec. IV we will draw our conclusions.

\chap{Form} {II. Z$^{\displaystyle\bf \prime}$ and new fermion effects.
           Formalism}

\nobreak
\bigskip
\nobreak
We will now review the formalism for describing the combined effects due
to the presence of a new neutral gauge boson, and of new fermions that
could be mixed with the known ones. We will follow closely the presentation
given in [\cite{fit6}] concentrating mainly on
the charged fermions sector and on flavor changing effects, and
we refer to [\cite{fit6}] for a more general discussion.
We will assume a low energy gauge group of the form
$\G\times U_1(1)$, where  $\G = SU(2)_L \times U(1)_Y \times
SU(3)_C$ is the usual SM gauge group.
Then in the gauge basis, the corresponding neutral current Lagrangian
reads

$$
-{\cal L}_{\rm NC}=eJ^\mu_{\rm em}A_\mu +
g_0 J_0^\mu Z_{0 \mu} + g_1 J_1^\mu Z_{1 \mu}.
\eq{2.1}
$$
\mn
In \req{2.1} $Z_0$ is the SM neutral gauge boson,
which couples with strength

$$
g_0 =(4\sqrt{2} G_F M^2_{Z_0})^{1/2}
\eq{2.2}
$$
to the usual combination of the neutral isospin and electromagnetic
currents
$$
J^\mu_0=J^\mu_3-s^2_W J^\mu_{\rm em}
\eq{2.3}
$$
where $s^2_W=\sin^2\theta_W$ is the weak mixing angle.
The new $Z_1$ corresponds to the additional $U_1(1)$ factor,
and couples to the new $J_1$ current with strength $g_1$.
In particular, if we assume that the $U_1(1)$ originates at low energy
from a GUT based on a $simple$ group,
then the $q_1(f)$ charges of the fermions are
fixed by the gauge group.
Normalizing the new generator $Q_1$
to the hypercharge axis $Y/2$ and assuming a similar renormalization
group evolution for the two abelian couplings $g_1$ and $g_{\scss Y}$
down to the electroweak scale, the
coupling strength of the new interaction can be written as
$$
g_1 \simeq g_0 s_W.
\eq{2.4}
$$
In general, after spontaneous symmetry breaking
the $Z_0$--$Z_1$ mass matrix turns out to be non diagonal.
The  $Z$ and $Z^\pr$ gauge bosons mass eigenstates
then correspond to two orthogonal combinations of
$Z_0$ and $Z_1$ that we will parametrize in term of
an angle $\phi$.
As a result, the currents that couple with strength $g_0$ to the physical
$Z$ and $Z^\prime$ are:
$$
\pmatrix{ J^\mu_Z \cr J^\mu_{Z^\prime} } =
\pmatrix{c_\phi&s_\phi\cr -s_\phi&c_\phi\cr}
\pmatrix{ J^\mu_0 \cr s_W J^\mu_1 }.
\eq{2.5}
$$
\mn
We see that the main effects of the presence of the new gauge boson
are the additional contribution to NC amplitudes,
described by the third term in the Lagrangian
\req{2.1}, and the mixing between the $J_0$ and $J_1$
currents in \req{2.5}.
We will not discuss additional indirect effects, as e.g. the shifts
induced by $Z_0$--$Z_1$ mixing in
the value of the weak mixing angle $s_W$  and in the overall coupling
strength $g_0$ when expressed as a function of the $Z$ mass
[\cite{fit6}-\cite{zp-new}],
since they are irrelevant for the present analysis.

We now discuss the form of the two currents $J_0$ and $J_1$,
and in particular the effects that could originate
from the presence of additional charged fermions.
We will henceforth denote as {\it new}, possible degrees
of freedom in addition to the standard 15 {\it known} fermions per generation.
Since no new fermions have been directly observed yet, the new
states must be rather heavy, and $m_{\rm new}\gtrsim M_Z/2$ can be
taken as the present model independent limit on the new masses.
Accordingly, we will denote the corresponding mass eigenstates as {\it
heavy}, while the known mass eigenstates will be labeled as {\it
light}.
In the presence of additional fermions, the light mass eigenstates
will correspond to superpositions of the known and new
states. Conservation of the electric and color charges forbids
a mixing between gauge eigenstates with different
$U(1)_{\rm em}$ and $SU(3)_{\rm c}$ quantum numbers,
and, in turn, this implies that
the electromagnetic and color currents of the
light mass eigenstates are not modified in the presence of the new
states.
However, the neutral isospin generator $T_3$ and the new generator $Q_1$ are
spontaneously broken, then a mixing between
gauge eigenstates with different $t_3$ and $q_1$ eigenvalues
is allowed, and as a result the couplings
of the light mass eigenstates to the $Z_0$ and $Z_1$ will be affected.

Since in the gauge currents chirality is conserved too,
it is convenient to group the fermions
with the same electric charge
and chirality $\alpha=L,R$
in a column vector of the known ($\cal K$) and new ($\cal N$) gauge
eigenstates  $\Psi^{o}_\alpha=(\Psi^o_{\K}, \Psi^o_{\N})_\alpha^T$.
The relation between the gauge eigenstates $\Psi^o_\alpha$
and the corresponding light and heavy mass eigenstates
$\Psi_\alpha=(\Psi_l,\Psi_h)_\alpha^T$
is then given by a unitary transformation
$$
\pmatrix{\Psi^o_{\K}\cr\Psi^o_{\N}}_\alpha = U_\alpha
\pmatrix{\Psi_l\cr\Psi_h}_\alpha \qquad{\rm where}\qquad
U_\alpha = \pmatrix{A &G\cr F & H}_\alpha ,
\qquad  \alpha=L,R.
\eq{2.6}
$$
\mn
The submatrices $A$ and $F$ describe the overlap of the light
eigenstates with the known and the new states respectively, and
from the unitarity of $U$ we have
$$
A^\dagger A+F^\dagger F=A A^\dagger +G G^\dagger =I.
\eq{2.7}
$$
Note that we have not introduced an extra index to
label the electric charge; nevertheless we will
treat $\Psi^o_\alpha$ and $\Psi_\alpha$
as vectors corresponding to a definite value of $q_{\rm em}$.

In terms of the fermion mass eigenstates
the neutral current  corresponding to a (broken)
generator ${\cal Q}$ reads
$$
J^\mu_{\cal Q} =\sum_{\alpha=L,R}
\bar\Psi_{\alpha} \gamma^\mu U^\dagger_\alpha {\cal Q}_\alpha
U_\alpha\Psi_{\alpha},
\eq{2.8}
$$
\mn
where ${\cal Q}_\alpha$ represents a generic diagonal
matrix of the charges for the chiral fermions.
Here we have to consider only the mixing effects in $J_3$
appearing in \req{2.3}
and in $J_1$,  since the term
proportional to $J_{\rm em}$ in $J_0$
is not modified by fermion mixing.
Hence in \req{2.8} ${\cal Q}=T_3,\, Q_1$ and
the elements of the corresponding matrices  are given by the
eigenvalues $t_3$ and $q_1$.

{}From \req{2.8} we see that if in one subspace of states with equal
electric charge and chirality
the matrix ${\cal Q}_\alpha$ is proportional to the identity,
then $U_\alpha^\dagger {\cal Q}_\alpha U_\alpha ={\cal Q}_\alpha$ and
for these fermions the corresponding current is not modified
in going to the base of the mass eigenstates.
In the SM for example, for a given electric
charge and chirality the $t_3$ eigenvalues of the fermions
are indeed the same,
and this implies in particular the absence (at the tree level)
of FCNC.                
In models with new fermions in contrast, the diagonal
matrices ${\cal Q}_\alpha$ have the general form
${\cal Q}_\alpha =
{\rm diag} ({\cal Q}^{\K}_\alpha,\, {\cal Q}^{\N}_\alpha )$
and do not commute with $U$.

To put in evidence the indirect effects
of fermion mixings in the couplings of the light mass eigenstates,
we now project \req{2.8} on the light components $\Psi_l$, obtaining
$$
J^\mu_{l{\cal Q}}
               =\sum_{\alpha=L,R}
\bar \Psi_{l\alpha} \gamma^\mu \left[
A^\dagger_\alpha {\cal Q}_\alpha^{\K}
A_\alpha +
F^\dagger_\alpha {\cal Q}^{\N}_\alpha F_\alpha \right]
\Psi_{l\alpha}.                   \eq{2.9}
$$
\mn
This equation is quite general, and describes
the effects of fermion mixings
in the neutral--currents of light--states for a wide class of models.
If the gauge group is generation independent, all the known states appearing
in one vector $\Psi^o_\alpha$ have the same eigenvalues with respect to
the generators of the gauge symmetry, and hence
we have ${\cal Q}^{\K}_\alpha= q_\alpha^{\K}I$ with
$q_\alpha^{\K}=t_3(f^{\K}_\alpha)$, $q_1(f^{\K}_\alpha)$.
The same is not true in general for the new states;
however, if we consider the particular case when
the mixing is with only one type of new fermions
with the same $q_\alpha^\N$ charges, then we have
${\cal Q}^{\N}_\alpha= q_\alpha^{\N}I$ as well.
Under these conditions, and by means of the unitarity relation \req{2.7},
\req{2.9} reduces to the simple form
$$
J^\mu_{l{\cal Q}}
               =\sum_{\alpha=L,R}
\bar \Psi_{l\alpha} \gamma^\mu \left[ q_\alpha^{\K} I +
(q_\alpha^{\N} - q_\alpha^{\K})
F^\dagger_\alpha F_\alpha \right]\Psi_{l\alpha}. \eq{2.10}
$$
\mn
We will restrict ourself to this equation, which is general
enough for describing the mixing with the additional charged fermions
in $\E$, and we refer to [\cite{fit6}] for a discussion of more general cases.

A few consideration are now in order.
In \req{2.10} the first term $q_\alpha^{\K}I $
inside the square brackets gives the couplings of a particular
light fermion in the absence of
mixing effects. The second term represents the
modifications due to fermion mixings.
The matrix $F^\dagger F$ appearing in this term
is in general not diagonal, and
while the magnitude of the diagonal elements will affect
the strength of the flavor diagonal couplings of the mass
eigenstates, the off diagonal terms will induce FCNC.
Clearly whenever the coefficient
$(q_\alpha^{\N} - q_\alpha^{\K})$ vanishes, the mixing effects are
absent.
When $t_3(f_\alpha^{\N}) \ne t_3(f_\alpha^{\K})$
the $J_0$ current is modified, and then the
existing low--energy and on--resonance NC data
as well as the current limits on rare processes,
can be used to constrain directly the corresponding
elements of $F^\dagger F$.
Model independent limits on the diagonal elements $(F^\dagger F)_{ii}$
affecting $Z_0$ interactions
were first given in [\cite{ll1}] and subsequently updated in
[\cite{fit}], and the most recent limits in the frame
of $\E$ models can be found in [\cite{fit6}].
Bounds on the off diagonal terms
$(F^\dagger F)_{i\ne j}$ have been given in [\cite{ll1}] as well.
All these limits turn out to be very stringent, usually at the level
of 1\% or better.
In contrast, if $t_3(f_\alpha^{\N}) = t_3(f_\alpha^{\K})$
the $J_0$ current is not modified, but since in general
we still have $q_1(f_\alpha^{\N}) \ne q_1(f_\alpha^{\K})$,
sizable effects could indeed be present in $J_1$.
We stress that the present experimental data
cannot be used to set limits on these mixings,
since they only affect a new hypothetical interaction,
and we can only rely on theoretical speculations to estimate
their magnitude.

According to these considerations it is clear that from a phenomenological
point of view
it is convenient to classify possible new fermions
in terms of their transformation properties under $SU(2)_L$.
Since we are only interested in fermions with conventional electric
charges, the new states must be singlets or doublets of weak--isospin.
A rather heterodox exception is that of a gauge triplet of
fermions [\cite{salati-trip}], but we will not consider this
possibility here.
According to the nomenclature in use [\cite{fit6},\cite{ll1},\cite{fit}],
we denote the particles with unconventional isospin assignments
(left--handed singlets or right--handed doublets) as {\it exotic} fermions.
All the standard fermions, as well as all the new states
that have conventional $SU(2)_L$ assignments, are referred to
as {\it ordinary}.
For example mirror fermions, having opposite $SU(2)_L$ assignments from
those of the known fermions, are exotic.
Sequential fermions are simply repetition of the new fermions.
They could be present in a complete new family or as components of
large fermion representations and are clearly classified as ordinary.
In this paper we will mainly concentrate on
vector multiplets of new fermions for which the L and R components
have the same $SU(2)_L$ transformation properties,
and hence always contain both ordinary and exotic states.
{}From the previous discussion, we see that while ordinary--exotic fermion
mixings are tightly constrained
due to the effects induced in the $J_0$ current,
no limits can be given for the ordinary--ordinary
mixings, since they affect only $J_1$.

This classification is also very convenient
for discussing  the possible form of the fermion mass matrices
and the expected magnitude of the mixings between the known and the
new fermions.
To give an example, let us introduce for each fermion family
a vector gauge singlet of new fermions
$({X^o_{\scss E}}_L,{X^o_{\scss O}}_R)_i$
($E$ = exotic, $O$ = ordinary, $i=1,2,3$)
with the same electric and color
charges than the  known fermions $({f^o_{\scss O}}_L,{f^o_{\scss
O}}_R)_i$. Then in the gauge eigenstate basis
the mass term reads
$$
{\cal L}_{\rm mass}=
{(\bar {f^o_{\scss O}}, \bar {X^o_{\scss E}})}_L \>
{\cal M} \> {f^o_{\scss O} \choose  X^o_{\scss O}}_R
+ {\rm h.c.}
\eq{2.11}
$$
where e.g. $f^o = (f^o_1,f^o_2,f^o_3)^T$ etc..
The non diagonal mass matrix ${\cal M}$ takes the form
$$
{\cal M} = \pmatrix{D &D^\pr\cr S^\pr &S},
\eq{2.12}
$$
\mn
where $D$ and $D^\pr$ are $3\times3$ matrices generated by
vacuum expectation values (vevs)
of doublets multiplied by Yukawa couplings, while
$S$ and $S^\pr$ are generated by vevs of singlets.
As a general rule, while the mass terms
which couple ordinary L-fermions to ordinary R-fermions
(or exotic L-fermions to exotic R-fermions)
arise from vevs of Higgs doublets, the entries
which couple ordinary fermions to the exotic ones are generated
by vevs of singlets.
Then in general Higgs singlets
are responsible for the large masses of new
heavy fermions in vector multiplets and,
in most cases, also contribute to the mass
of the new heavy gauge boson; hence it is natural to
assume $S,S^\pr \sim \Lambda \gg D,D^\pr$.

The diagonal mass matrix M is obtained via a biunitary transformation
acting on the L and R sectors:
$$
\def\oe{\scss{O{\rm-}E}}
\def\oo{\scss{O{\rm-}O}}
\eqalign{
{\rm M}^2
&=U_L^{\oe}({\cal M}{\cal M}^\dagger) {U_L^{\oe}}^\dagger \cr
&=U_R^{\oo}({\cal M}^\dagger{\cal M}) {U_R^{\oo}}^\dagger.  \cr
} \eq{2.13}
$$
Since $D/\Lambda$, $D^\pr/\Lambda\sim \varepsilon \ll 1\,$,
the order of magnitude of the different entries in
${\cal M}{\cal M}^\dagger $ and ${\cal M}^\dagger {\cal M}$ is

$$
\eqlabel{2.14} \eqlabel{2.15}
\eqalignno{
{\cal M}{\cal M}^\dagger &
\sim \Lambda^2 \pmatrix{\varepsilon^2 &\varepsilon \cr
              \varepsilon   &1       \cr}     &\req{2.14}\cr
{\cal M}^\dagger{\cal M} &
\sim \phantom{\Lambda}\pmatrix{\Lambda^2 &\Lambda^2 \cr
              \Lambda^2 &\Lambda^2 \cr}.             &\req{2.15}\cr }
$$

Given the form of
${\cal M}{\cal M}^\dagger$ in  \req{2.14}
and keeping in mind the expression \req{2.6} for
the matrices  $U$,
we see that for the matrix
describing the ordinary--exotic mixings in \req{2.13}
it is natural to expect that the
submatrices $F$ and $G$
would acquire an overall suppression factor
$\varepsilon$, of the order of the ratio of the light to heavy mass
scale. In contrast, since all the entries in \req{2.15} are of the
same order of magnitude, such a suppression is not present for the
ordinary-ordinary $F$ and $G$ mixing terms.
Now, since it is precisely $F^\dagger F$
in \req{2.10} which affects the flavor
diagonal couplings and also induces  FCNC,
the suppression of the ordinary--exotic mixings
explains in a natural way the non--observations of these effects
in the $Z_0$ interactions.
On the other hand for the ordinary--ordinary mixings
there is no reason to expect the elements of
$F^\dagger F$ to be particularly small,
and accordingly flavor changing processes can be expected to occur
at a sizeable rate in $Z_1$ interactions.

Written explicitly, the flavor diagonal chiral couplings of a
light $f$ fermion to the $Z_0$ and $Z_1$ gauge bosons are
$$
\eqalign{
\varepsilon_{0\alpha}^f &= t_3(f_\alpha) - s^2_W q_{\rm em}(f)
\ + \ [t_3(f^{\N}_\alpha) - t_3(f^{\K}_\alpha)] \,
(F_\alpha^\dagger F_\alpha)_{ff}
\cr
\varepsilon_{1\alpha}^f &= q_1(f_\alpha) \ + \
[q_1(f^{\N}_\alpha) - q_1(f^{\K}_\alpha)] \,
(F_\alpha^\dagger F_\alpha)_{ff},
\qquad\qquad\qquad \alpha=L,R.  \cr
}
\eq{2.16}
$$
while the $f_i\,f_j$ flavor changing couplings read
$$
\eqalign{
\kappa_{0\alpha}^{ij} &= [t_3(f^{\N}_\alpha) - t_3(f^{\K}_\alpha)]
\, (F_\alpha^\dagger F_\alpha)_{ij}
\cr
\kappa_{1\alpha}^{ij} &= [q_1(f^{\N}_\alpha) - q_1(f^{\K}_\alpha)]
\, (F_\alpha^\dagger F_\alpha)_{ij},
\qquad\qquad\qquad \alpha=L,R.  \cr
}
\eq{2.17}
$$
The corresponding couplings
to the physical $Z$ and $Z^\prime$ bosons, that we will denote as
$\varepsilon_\alpha^f$, $\varepsilon_\alpha^{\pr\,f}$
and
$\kappa^{ij}_\alpha$, $\kappa_\alpha^{\pr\, ij}$,
can be readily obtained via the transformation \req{2.5}.
For the flavor changing couplings we have for example
$$
\eqalign{
\kappa^{ij}_\alpha &=c_\phi\, \kappa^{ij}_{0 \alpha} +
s_\phi s_w \, \kappa^{ij}_{1 \alpha}          \cr
\kappa_\alpha^{\pr\, ij} &=-s_\phi \, \kappa^{ij}_{0 \alpha}
 + c_\phi s_w \, \kappa^{ij}_{1 \alpha}
\qquad \qquad \qquad \qquad  \alpha=L,R, \cr }
\eq{2.18}
$$
\mn
and analogous expressions hold for
$\varepsilon_\alpha^f$ and $\varepsilon_\alpha^{\pr\,f}$ too.
In terms of the flavor non--diagonal couplings \req{2.18},
the FCNC Lagrangian for the light $f^i$ and $f^j$ fermions
in the mass eigenstate basis finally reads
$$
-{\cal L}_{\rm FC}^{ij}=
g_0 \sum_{\alpha=L,R} \left(
\bar{f^i_\alpha}\gamma^\mu \, \kappa^{ij}_\alpha \, f^j_\alpha \, Z_\mu
\ + \
\bar{f^i_\alpha} \gamma^\mu \, \kappa^{\pr ij}_\alpha \,
f^j_\alpha Z^\pr_\mu\right).
\eq{2.19}
$$

{}From the first equation in \req{2.18} we see that
even in the case when only ordinary--ordinary mixing effects
are present and hence $\kappa^{ij}_{0 \alpha} = 0$,
the $Z$ boson can still mediate flavor changing transitions,
suppressed now by a factor proportional to the $Z_0$--$Z_1$
mixing [\cite{zp-fc}].
However, for several models the existing limits on $\phi$ are rather
stringent: $|\phi| \lesssim 0.02$ [\cite{fit6}-\cite{zp-new}] and
then we can expect that if the $Z^\pr$ is not too heavy, FCNC
processes would be mainly induced by $Z^\pr$ exchange.

\chap{Constr}III. Constraints on E$_{\displaystyle\bf 6}$ models from
\bfmu $\ \rightarrow$ \bfe\bfe\bfe.

In the previous section
we have shown that in the presence of new charged fermions
the new neutral current $J_1$ could induce sizeable
flavor changing transitions either via $Z^\pr$
interactions, or as a consequence of a non--vanishing $Z_0$--$Z_1$ mixing.
This kind of new  physics
would manifest itself in affecting the rates for several
processes which are forbidden or highly suppressed in the SM.
In the quark sector it could enhance the branchings for the
leptonic decays of mesons like $K^0$, $D^0$, $B^0\rightarrow \ell^+\ell^-$,
and it would also affect the neutral meson mixings and mass
differences.
In the lepton sector it would induce
several LFV neutrinoless $\tau$ decay modes like
$\tau \rightarrow
eee$, $\mu\mu\mu$, $\mu ee$, $e\mu\mu$, $\mu \pi$, $\mu \rho$
which are all constrained at the level of
$\lesssim$ few$\times 10^{-5}$ [\cite{pdg92}],
but in particular it would also give rise to the decay
$\mu \rightarrow eee $ for which the existing
limit [\cite{sindrum}] is
much more stringent:

$$
Br(\mu^+\rightarrow e^+e^+e^-)<1.0\cdot 10^{-12} \hskip 1truecm
({\rm at\quad} 90\% \, c.l.\,) .
\eq{3.1}
$$
\mn
Given the LFV Lagrangian \req{2.19}, the expression for this
decay rate relative to the charged current decay $\mu \rightarrow
\nu e\bar\nu $ is

\def\e{{\varepsilon}}
\def\ep{{\varepsilon^\prime}}
\def\k{{\kappa}}
\def\kp{{\kappa^\prime}}
$$
\eqlabel{3.2}
\eqalignno{
{Br(\mu\rightarrow eee) \o Br(\mu\rightarrow \nu e\bar\nu)} &=
2\Bigl[3(\e_R^2+\e_L^2)(\k_R^2+\k_L^2)+
(\e_R^2-\e_L^2)(\k_R^2-\k_L^2)\Bigr]  +                 \cr
2({M^2_Z\o M^2_{Z^\pr}})^2&\Bigl[3(\ep_R^2+\ep_L^2)(\kp_R^2+\kp_L^2)+
(\ep_R^2-\ep_L^2)(\kp_R^2-\kp_L^2)\Bigr] +         &\req{3.2}     \cr
4{M^2_Z\o M^2_{Z^\pr}}
\bigl[3(\e_R\ep_R  &+\e_L\ep_L)(\k_R\kp_R+\k_L\kp_L)+
(\e_R\ep_R-\e_L\ep_L)(\k_R\kp_R-\k_L\kp_L)\bigr]      \cr}
$$
where for $\k_{R,L}$ and $\kp_{R,L}$ defined in \req{2.18}
we have dropped the indices $i=e$ and $j=\mu$, and
$\e_{R,L}$ and $\ep_{R,L}$ refer to the electron couplings.

As a first result, by confronting  \req{3.1} and \req{3.2}
we can derive the limits on the $Z\mu e$ LFV vertices.
Assuming  that the $Z^\pr$ is completely decoupled from the
low energy physics ($M_{Z^\pr}\rightarrow \infty$ and $\phi
\rightarrow 0$) and taking $s^2_W=0.23$, we obtain:
$$
\eqalign{
|\k^{e\mu}_L| &< 1.1\cdot 10^{-6}  \cr
|\k^{e\mu}_R| &< 1.2\cdot 10^{-6}.  \cr}
\eq{3.3}
$$
As we have discussed, if these couplings do originate from
some kind of mixing of the electron and muon
with heavy exotic leptons,
we expect them to be strongly suppressed, e.g. by a
factor of the order $m^2_\mu/M^2_Z\sim 10^{-6}$ or smaller, and
then we see that the existence of FCNC induced by
ordinary--exotic mixings does not conflict with the stringent limits in
\req{3.3}.

Now, in order to study the effects of the $Z^\pr$ flavor changing vertices,
we first need to fix the $q_1$ charges of the fermions, which
in turn determine the coefficient of the flavor changing
mixings.
This can be done by choosing a specific GUT,
and we will carry out our analysis in the frame of E$_6$.
Since E$_6$ has rank 6, while the
SM gauge group $\G$ has rank 4, the breaking of $\E$ to the SM will
lead to extra $Z^\pr$s. We will consider the possibility that
either $\E$ breaks directly to rank 5, or that one of the two extra
$Z^\pr$s is heavy enough so that its effects on the low energy physics
are negligible, and in these cases the formalism developed in
the previous section can be straightforwardly applied.
We will choose the  embedding of $\G$ into $\E$ trough the
maximal subalgebra chain
$
\E \rightarrow \ U(1)_{\psi} \times SO(10)
\rightarrow \ U(1)_{\chi} \times SU(5)\rightarrow \G
$,
then an effective extra $U_1(1)$ could arise
at low energy as a combination of the $U(1)_\psi$ and
$U(1)_\chi$ factors.
We will parametrize this combination in terms of an
angle $\beta$, and this will define an entire
class of $Z^\pr$ models in which each fermion $f$ is coupled
to the new boson through the effective charge
$$
q_1(f)=q_\psi(f) \sin\beta + q_\chi(f) \cos\beta .
\eq{3.4}
$$
Particular cases that are commonly studied in the literature
[\cite{fit6}--\cite{zp-new},\cite{rizzo-e6}]
correspond to $\sin\beta=-\sqrt{5/8}$, 0, 1
and are respectively denoted $Z_\eta$, $Z_\chi$ and $Z_\psi$ models.
$Z_\psi$ occurs in $\E\to$ SO(10), while
$Z_\eta$ occurs in superstring models when
$\E$ directly breaks down to rank 5.
As we will see this model plays a peculiar role in the present analysis,
since it evades completely the kind of constraints that we are investigating.
Finally, a $Z_\chi$ boson occurs in SO(10)$\to$ SU(5) and
couples to the conventional fermions in the same way
than the $Z^\pr$ present in SO(10) GUTs,
however since SO(10) does not contain additional charged fermions, the
kind of flavor changing effects that we are studying here is absent.
In contrast, new charged quarks and leptons are present
in $\E$. In the GUTs based on this gauge group
the fermions are assigned to the
fundamental {\bf 27} representation that contains,
beyond the standard 15 degrees of freedom,
12 additional states for each generation,
among which we have a vector doublet of new leptons
$(N\> E^-)_L^T$, $(E^+ \> N^c)_L^T$ on which we will now concentrate.

The chiral couplings of the leptons to the $Z_1$ and the coefficient
of the LFV term, are determined by the $q_\psi$ and $q_\chi$
charges of the new and known states, which are
$$
\eqlabel{3.8}
\eqalignno{
q_\psi(E_L)=-q_\psi(E_R)=-{1\o 3}\sqrt{5\o 2} \hskip 1.2truecm
&q_\chi(E_L)=q_\chi(E_R)=-{1\o 3}\sqrt{3\o 2}   & \cr
q_\psi(e_L)=-q_\psi(e_R)={1\o 6}\sqrt{5\o 2}  \hskip 1.8truecm
&q_\chi(e_L)=3q_\chi(e_R)={1\o 2}\sqrt{3\o 2}.   & \req{3.8} \cr
}
$$
With respect to the $SU(2)_L$ transformation properties, the $E^+_L$ heavy
leptons are exotic, and than the mixing of their CP conjugate states
$E^-_R$ with $e_R$, $\mu_R$ and $\tau_R$ are constrained by $Z_0$
interactions.
{}From \req{3.3} we have for example
$$
(F_R^\dagger F_R)_{e\mu} < 2.4\cdot 10^{-6},
\eq{3.5}
$$
while the 90\% c.l. limits on the flavor diagonal mixings given in
[\cite{fit6}] are respectively
$$
\eqlabel{3.6}
\eqlabel{3.7}
\eqalignno{
(F_R^\dagger F_R)_{ee} &< 1.3\cdot 10^{-2}     & \req{3.6} \cr
(F_R^\dagger F_R)_{\mu\mu} &< 1.1\cdot 10^{-2}. & \req{3.7} \cr
}
$$
\mn
Due to the tight bound \req{3.5} it is reasonable to
neglect the LFV couplings in the R-sector and
(conservatively) set $\k_R^{e\mu}=\kp_R^{e\mu}=0$ in \req{3.2}.
According to \req{3.6}, it is also justified to
neglect the effects of the fermion mixings in the
flavor diagonal couplings of the R-electrons.

In contrast, the $E^-_L$ leptons are ordinary, and no bounds exist
on their mixing with the light leptons.
We will still neglect the diagonal term
$(F_L^\dagger F_L)_{ee}$ since it is reasonable to
expect that if this term were so large as to spoil the approximation
$\e_{1\alpha}^f\simeq q_1(f_\alpha)$
the value of the off-diagonal term
$(F_L^\dagger F_L)_{e\mu}$ would also be large,
possibly leading to even stronger limits than the ones
derived here.

Due to the approximations made, for each value of the parameter
$\beta$ in \req{3.4}
the branching ratio \req{3.2} depends on the values of
$M_Z^\pr$, $\phi$ and ${\cal F}_{e\mu}\equiv(F_L^\dagger F_L)_{e\mu}$.
However, it is easy to see that since the gauge boson mixing effects in the
diagonal electron couplings are in any case very small,
being $|\phi| \lesssim 0.02$ [\cite{fit6}-\cite{zp-new}],
the relevant variables are actually only two, namely
${\cal F}_{e\mu} \cdot (M_Z^2/M_{Z^\pr}^2)$ and
${\cal F}_{e\mu} \cdot \phi$.
Moreover once the Higgs sector of the model
is specified, $M_{Z^\pr}$ and $\phi$ are
no more independent quantities.
For example an approximate relation that holds for
small mixings and when $M_{Z^\pr}$ ($\gg M_Z$)
originates from a large Higgs singlet vev [\cite{l-luo}] reads
$$
\phi \simeq - {M_Z^2\o M^2_{Z^\pr}} s_W
{\sum_i t_3^i q_1^i |\langle\phi^i\rangle|^2
\o
\sum_i {t_3^i}^2 |\langle\phi^i\rangle|^2 },
\eq{3.9}
$$
and in this case the branching ratio \req{3.2}
is in practice only a function of \break
${\cal F}_{e\mu}\cdot(M^2_Z/ M_{Z^\pr}^2)$.

As in the SM for the Cabibbo-Kobayashi-Maskawa (CKM) matrix,
also in $\E$ we don't have a clue for predicting
the values of the fermion mixing parameters.
Without attempting to push too far an analogy between
the mixings we are interested in
and the CKM matrix,
we will merely note that both these cases involve mixings among ordinary
fermions, and that we do not expect in the present case any
additional suppression factor. We also note that
all the CKM matrix elements are $> 10^{-3}$ and that in particular
the mixing between the first and the second generation is rather
large.
We will then assume that the LFV term ${\cal F}_{e\mu}$
lies in the range $10^{-2}$--$10^{-4}$. Under this assumption
the presence of a too light $Z^\pr$
as well as a too large amount of $Z_0$--$Z_1$ mixing will
clearly conflict with the limit \req{3.1}.
The bounds that can be derived in this way are indeed very strong,
but obviously they cannot replace the
direct [\cite{zp-direct}] or indirect [\cite{fit6}-\cite{zp-new}]
limits on the $Z^\pr$
parameters, since for very small values of the
LFV $Z^\pr$ couplings
(${\cal F}_{e\mu} \lesssim 10^{-6}$)
they would in fact be weaker.
We will nevertheless present our
constraints in the form of numerical limits on $M_{Z^\pr}$ and $\phi$,
since, in doing so, the strength of the arguments that have been
discussed here is put in clear evidence.

Our results are collected in Figs. 1 and 2.
Figure 1 shows the bounds on $M_Z^\pr$,
the thick solid line depicts the bounds obtained
for ${\cal F}_{e\mu} = 10^{-2}$ and by setting the
gauge boson mixing angle $\phi$ to zero. The decay $\mu\rightarrow
eee$ is due only to $Z^\pr$ exchange in this case.
The limits for different values of
${\cal F}_{e\mu}$ can be red off this line as well,
by assuming for the vertical axis units of
of GeV$\cdot [100{\cal F}_{e\mu}]^{1\o 2}$.
The thick dashed line, drawn here for convenience, shows
the bounds corresponding to $\phi=0$ and
${\cal F}_{e\mu} = 10^{-3}$, vertical units are again
in GeV in this case.
We see that for ${\cal F}_{e\mu}> 10^{-3}$
a $Z^\pr$ below 1 TeV would be excluded for most of the values
of $\beta$. Also, it is clear that
${\cal F}_{e\mu}\simeq  10^{-4}$ still leads to significative
bounds, being $M_{Z^\pr}$ constrained to values $\gtrsim 400$GeV
for large part of the $\sin\beta$ axis.

To study the possible effects on these results
 of a non vanishing
mixing angle $\phi$, i.e. when both the $Z^\pr$ and $Z$ bosons
contribute to the decay, we have used \req{3.9}
assuming two doublets of Higgs fields $h_{N^c}$ and $h_N$
with vevs $\bar v$  and $v$.
Since $\bar v$  and $v$ give mass respectively to the
$t$ and $b$ quarks, $\sigma \equiv {\bar v}^2/v^2 > 1$ is
theoretically preferred.
The bounds on $M_{Z^\pr}$ obtained by allowing
for a $Z_0$--$Z_1$ mixing consistent with this minimal Higgs
sector are shown Fig. 1 by the dotted and dot-dashed lines,
which correspond to $\sigma = 1$ and $\infty$ respectively.
It is apparent that by allowing for a non vanishing
value of $\phi$, the limits on the $Z^\pr$ mass are only slightly affected.

Figure 2 depicts the constraints on the $Z_0$--$Z_1$ mixing angle
$\phi$. The solid line shows the bounds obtained by
assuming ${\cal F}_{e\mu} = 10^{-2}$ and taking
the limit $M_{Z^\pr} \rightarrow \infty$.
In this case the decay $\mu \rightarrow eee$ is mediated only by
the $Z$ boson, and is due to the mixing between the $Z_0$ and the $Z_1$.
The dotted line shows the bounds for
${\cal F}_{e\mu} = 10^{-3}$ in the same limit.
The limits for different choices of ${\cal F}_{e\mu}$
are easily obtained from the solid [dashed] lines by rescaling
the vertical units by
$(10^2 {\cal F}_{e\mu})^{-1}$ \ [$(10^3 {\cal F}_{e\mu})^{-1}$].

The dotted ($\sigma=1$) and dot--dashed lines ($\sigma=\infty$)
enclose the regions of the limits obtained assuming a minimal Higgs
sector. In this case the value of $M_{Z^\pr}$ is
finite and consistent, according to \req{3.9},
with the values of $\phi$ at the bound.
We see that with this additional constraint
significant limits are found
for ${\cal F}_{e\mu} = 10^{-4}$ as well.

{}From Fig. 2 it is apparent that for a minimal Higgs sector
the limits on $\phi$ are significantly tighter
than in the $M_{Z^\pr} \rightarrow \infty$ limit, showing that
\req{3.1} in first place gives direct constraints on
the $Z^\pr$ mass, while the bounds on the $Z_0$--$Z_1$ mixing
obtained independently of $M_{Z^\pr}$ are weaker.
We note that this
behaviour is opposite to what is encountered in deriving
limits on the $Z^\pr$ parameters from precise electroweak data
[\cite{fit6}-\cite{zp-new}],
where in fact the best bounds on the $Z^\pr$ mass are obtained
from the tight limits on $\phi$ implied by the LEP measurements.

{}From Figs. 1 and 2 it is apparent that for the $\eta$ model,
corresponding to $\sin \beta=-\sqrt{5/8}$, both the
$Z^\pr$ mass and the $Z_0$--$Z_1$ mixing angle are not
constrained by the present analysis. This is due to the fact
that in this model, for the
ordinary--ordinary fermion mixings  besides
$t_3^\K=t_3^\N$ we also have $q_\eta^\K=q_\eta^\N$,
implying that both the coefficients of the $F_L^\dagger F_L$ term
in the $J_3$ and in the $J_1$ currents vanish.
This happens also in the quark sector and in the neutral sector,
hence the unsuppressed flavor
changing vertices are completely absent for the $Z_\eta$ boson,
and this insures that
besides the decay $\mu \rightarrow eee$ no other
processes can be found for implementing this kind of
constraints for the $\eta$ model.
The reason for this can be understood by considering
the decomposition $\E \rightarrow SU(6)\times SU(2)_I$,
where $SU(6)$ contains the SM group, while the
$SU(2)_I$ is ``inert" in the sense that $I_{3I}$ does not
contribute to the $Q_{\rm em}$ generator
[\cite{rosner86},\cite{slansky}].
$I_{3I}$ corresponds to
$\beta=\arctan\sqrt{3/5}$ in \req{3.4} and is orthogonal to $Q_\eta$
[$\beta=\arctan(-\sqrt{5/3})$]
which is then contained in the $SU(6)$ factor as well.
The fermions in the {\bf 27} of $\E$ with the same SM quantum numbers
($q_{\rm em}$, $t_3$, color) form multiplets (singlets and doublets) of
$SU(2)_I$ and clearly these multiplets also carry definite values of
the $Q_\eta$ charge. All the ordinary fermions with the same
color and electric charges, being members of the same $SU(2)_I$
multiplet,  have also the same $q_\eta$, and this
implies the absence of both the diagonal and the flavor changing
ordinary--ordinary mixing effects.
In contrast ordinary--exotic fermion mixing could still give rise to FCNC
also in the $\eta$ model,
but as already discussed the corresponding transitions are expected to be
largely suppressed, and do not imply any useful constraint.

According to this discussion, if an additional $Z^\pr$
with a mass of a few hundreds GeV is found together with new fermions
that could fit in the {\bf 27} of $\E$,
the observed absence of unsuppressed FCNC would suggest that
it could most probably be a $Z_\eta$.

\chap{Conclusions}   {IV. Conclusions}

We have carried out an analysis of models that predict a new
neutral gauge boson and new charged fermions from the point of view
of FCNC processes.
We have argued that in most of these models
unsuppressed flavor changing couplings of the light fermions to the
new $Z^\pr$
can be present as a consequence of a mixing between the known
and the new charged fermions.
By assuming that these flavor changing vertices should not be
unnaturally small,
we have inferred that the observed suppression of FCNC processes
can still be explained in a natural way if
the new gauge boson is sufficiently heavy
and almost unmixed with the standard $Z$.
Also we have attempted a semi-quantitative analysis
of this kind of new physics in the frame of $\E$ models,
by confronting the theoretical expectations for the LFV effects
with the extremely
stringent limits on the $\mu \rightarrow eee$ decay mode.
Our conclusions are that the existence of $Z^\pr$ bosons from $\E$
much lighter than 1 TeV is unlikely, with the noticeable
exception of the superstring inspired $\eta$ model which
evades completely our constraints.
At the same time, our analysis suggests that
the observation of FCNC processes at rates larger
than the SM predictions could be interpreted as a hint for the
simultaneous presence of additional gauge bosons and new charged fermions.
Indeed these new states could manifest themselves indirectly via this kind of
flavor changing effects well before they are directly produced.

\vfill\eject


\null
\baselineskip 8pt
\centerline{\title References}
\vskip .8truecm

\biblitem{zp-fc}
T.K. Kuo and N. Nakagawa, \prd{32} 306 (1985).\par

\biblitem{e6-fcloop}
J. Bernabeu \ea, \plb{187} 303 (1987). \par

\biblitem{e6-fcrizzo}
G.Eilam and T.G. Rizzo, \plb{188} 91 (1987). \par

\biblitem{salati-trip}
B.W. Lee, \prd{6} 1188 (1972); \hbup
J. Prentki and B. Zumino, \npb{47} 99 (1972); \hbup
P. Salati, \plb{253} 173 (1991). \par

\biblitem{zp-direct}
CDF Collaboration, F. Abe \ea, \prl{67} 2609 (1991); \ib {\bf 68},
1463 (1992). \par

\biblitem{fit6}
E. Nardi, E. Roulet and D. Tommasini,
Report No. FERMILAB PUB-92/127-A 1992,
to appear in \prd{46} (1 october 1992). \par

\biblitem{l-luo}
P. Langacker and M. Luo, \prd{45} 278 (1992). \par

\biblitem{zp-new}
J. Layssac, F.M. Renard and C. Verzegnassi, \zpc{53} 97 (1992); \hbup
M.C. Gonzalez Garc\'\i a and J.W.F. Valle; \plb{259} 365 (1991); \hbup
G. Altarelli et al., \plb{263} 459 (1991); \hbup
F.del Aguila, J.M. Moreno and M. Quir\'os, \npb{361} 45 (1991); \hbup
F. del Aguila, W. Hollik, J.M. Moreno and M. Quir\'os, \ib {\bf B372}
 3 (1992). \par

\biblitem{rizzo-direct}
P. Langacker and M. Luo, in ref. [\cite{zp-new}];  \hbup
T.G. Rizzo; talk given at the 15$^{th}$ Johns Hopkins
Workshop on Current Problems in Particle Theory, Baltimore, MD, August
26-28 1991; ANL-HEP-CP-91-85. \par

\biblitem{fit}
E. Nardi, E. Roulet and D. Tommasini.
Report no. FERMILAB Pub-91/207-A,
to appear in \npb{}; \hbup
E. Nardi, E. Roulet and D. Tommasini;
Contributed to Int. Workshop on Electroweak Physics beyond the Standard
Model, Valencia, Spain, Oct 2-5, 1991
[Fermilab Report No. Conf-91/335-A 1991 (unpublished)]. \par

\biblitem{ll1}
P. Langacker and D. London, \prd{38} (1988) 886.\par

\biblitem{ll2}
P. Langacker and D. London, \prd{38} (1988) 907.\par

\biblitem{slansky}
For a review on the group E$_6$, see R. Slansky; \prep{79} (1981) 1. \par

\biblitem{pdg92}
Particle Data Group, J.J. Hern\'andez {\it et al.},
\prd {\bf 45} Part. II (1992). \par

\biblitem{sindrum}
SINDRUM collaboration, U. Bellgardt \ea, \npb{299}
(1988) 1. \par

\biblitem{rizzo-e6}
J.L. Hewett and T.G. Rizzo, \prep{183} (1989) 195, and references
therein. \par

\biblitem{rosner86}
D. London and J.L. Rosner, \prd{34} (1986) 1530. \par

\interlinea


\insertbibliografia
\vfill\eject

\centerline{\lltitle Figure captions}
\bigskip
\baselineskip 14pt

\medskip\noindent
Fig. 1:
Limits on $M_{Z^\pr}$ for a general $\E$ neutral gauge boson, as a
function of $\sin\beta$ and for different values of
the lepton flavor violating term ${\cal F}_{e\mu}\equiv(F_L^\dagger
F_L)_{e\mu}$.
The thick solid line is obtained by setting
the $Z_0$--$Z_1$ mixing angle $\phi$ to zero, and assuming
${\cal F}_{e\mu}=10^{-2}$.
The limits for different values of ${\cal F}_{e\mu}$
can be read off this line by assuming for the vertical axis units of
GeV$\cdot (100 {\cal F}_{e\mu})^{1\o 2}$. The thick dashed line
depicts the limits corresponding to ${\cal
F}_{e\mu}=10^{-3}$ (vertical units in GeV). The bounds obtained by
allowing for a non-vanishing $Z_0$--$Z_1$ mixing, consistent
with the values of $M_Z^\pr$ when a minimal Higgs sector is assumed, are also
shown. The dotted lines correspond to equal vevs of the two Higgs
doublets present in the model, {\it i.e.}
$\sigma\equiv {\bar v}/v=1$ while the dot--dashed lines
correspond to $\sigma=\infty$.

\bs\bs\bs
\medskip\noindent
Fig. 2:
Limits on the $Z_0$--$Z_1$ mixing angle $\phi$ for a general $Z_1$
from $\E$, as a function of $\sin\beta$ and for different values of
the lepton flavor violating term ${\cal F}_{e\mu}\equiv(F_L^\dagger
F_L)_{e\mu}$. The thick solid and dashed lines
are obtained in the
limit $M_{Z^\pr}\rightarrow \infty$ assuming
${\cal F}_{e\mu}=10^{-2}$ and
${\cal F}_{e\mu}=10^{-3}$ respectively. Limits for different values of
${\cal F}_{e\mu}$ can be obtained from these lines by rescaling the
vertical units by $(10^2 {\cal F}_{e\mu})^{-1}$ and $(10^3 {\cal
F}_{e\mu})^{-1}$ respectively. The dotted $(\sigma=1)$ and dot-dashed
$(\sigma=\infty)$ lines show the limits obtained for a finite $Z^\pr$
mass and assuming a minimal Higgs sector. In this case the bounds are
tighter and are essentially determined by the corresponding limits on
$M_{Z^\pr}$ through eq. \req{3.9}.  Also the limits corresponding to
${\cal F}_{e\mu}=10^{-4}$ are shown in this case.

\vfill\eject
\bye